# Achieving sub-1 Ohm-mm Non-Recess S/D Contact Resistance in GaN HEMTs Utilizing Simple CMOS Compatible La/Ti/Al/Ti Metal Contacts

Xinpeng Lin, Yumeng Zhu, Yongle Qi, Guangnan Zhou, Wenmao Li, Yang Jiang, Jian Zhang, Robert Sokolovskj, Yulong Jiang, Guangrui (Maggie) Xia, Mengyuan Hua and Hongyu Yu

***Abstract*—In this paper, we report the use of lanthanum (La) in S/D contacts of GaN HEMTs, achieving 0.97 Ohm-mm contact resistance without S/D recess. The HEMTs show well-behaved electrical characteristics and satisfactory reliability. Our studies show that La, a CMOS compatible metal, is promising to lower GaN HEMT S/D contact resistance. Spherical-shaped high-La regions can be observed near the surface after annealing. La diffuses into the AlGaN layer, and the overlap of La and Al peaks is significantly increased compared with that before the annealing. On the one hand, La's low work function (3.5 eV) is beneficial for reducing the barrier between the metals and GaN. The Ohmic contact formation mechanism involved was shown to be different from conventional Ti/Al films.**

## I. Introduction

As an excellent power device material, GaN has a wide band gap, a large critical breakdown voltage, and an excellent thermal conductivity. These properties enable GaN devices to be used at extreme high power densities [1]. Making Source/Drain (S/D) Ohmic contacts and enhancement-mode (E-mode) device operation with a threshold voltage $V_{th} > 3$ V are among the biggest challenges for AlGaN/GaN based high-electron-mobility transistors (HEMTs) [2].

Ohmic contacts are fundamental building blocks of power devices [3]. As Ohmic contacts link the devices to external circuitry, their resistance needs to be negligible with respect to that of the channel, in order to minimize the device specific on-resistance ($R_{on}$) and, hence, the power losses of the system. Gold (Au)-based S/D contacts are commonly used with good performance. However, Au is not a CMOS compatible metal, as it is a fast-diffusing contaminant in Si that deteriorates the minority carrier lifetime [4]. Therefore, GaN metallization schemes need to be Au-free to integrate with Si-based CMOS. In addition, Au-free CMOS compatible Ohmic contacts can greatly reduce the production cost. Studies on CMOS compatible Ohmic contacts to n-type GaN have been focused on metals with a low work function, such as Ti (4.33 eV [5]) or Al (4.28 eV [5]). However, the use of these metals requires recessing process in S/D areas before contact metal deposition which is hard to control [5], or else 1 Ohm-mm contact resistance is hard to be obtained.

In the first part of this paper, we report the use of lanthanum (La), which is a CMOS compatible metal with a much lower work function (3.5 eV [5]), in n-GaN HEMT S/D Ohmic contacts. The La-based contacts achieved around 0.97 Ohm-mm without recessing the S/D areas. In the second part of this paper, we will discuss the mechanisms of the La-based low $R_C$ contacts. Some difference exists in the mechanisms compared to conventional Ti/Al-based contacts, where Ti forms a TiN alloy with N in GaN after annealing [6].

## II. La-based Source/Drain Ohmic Contact

### A. *Fabrication Process La-based Ohmic Contact*

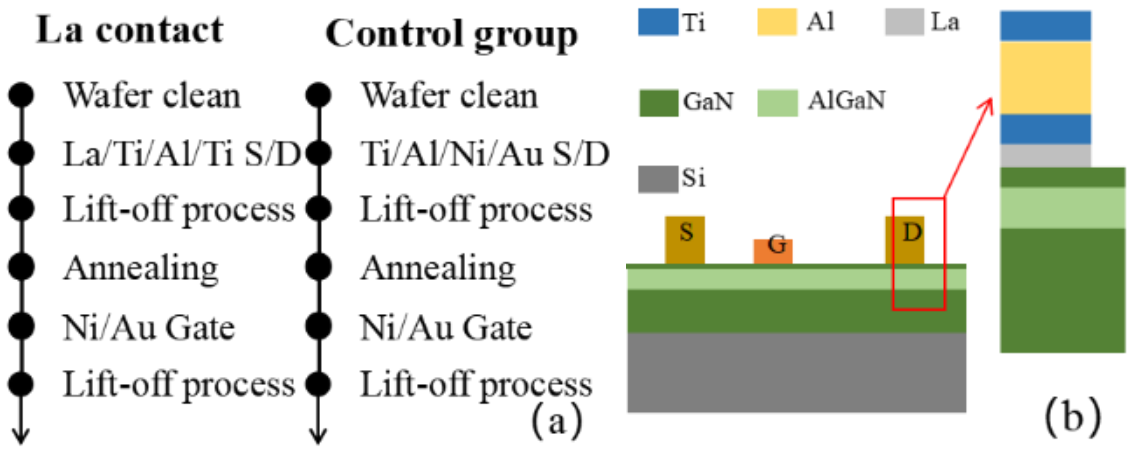

Fig. 1. (a) Process flows of the HEMTs with the La-based contacts and the control group with Au-based contacts. (b) Schematic diagram of the HEMTs and the La-based S/D metal stacks.

HEMTs and contact structures in this work have undoped GaN-cap (2 nm)/$Al_{0.25}Ga_{0.75}N$ (20 nm)/GaN layers grown by metalorganic chemical vapor deposition (MOCVD) on Si (111)

This work is sponsored by the project of “Research of low cost fabrication of GaN power devices and system integration” research fund (Grant No: JCYJ20160226192639004), the project of "Research of AlGaN HEMT MEMS sensor for work in extreme environment" (Grant No: JCYJ20170412153356899) and “Research of the reliability mechanism and circuit simulation of GaN HEMT” (Grant No:2017A050506002).

Xin-Peng Lin, Yumeng Zhu, Yongle Qi, Guangnan Zhou, Wenmao Li,Yang Jiang, Jian Zhang, Robert Sokolovskj, Mengyuan Hua, Lingli Jiang and Hongyu Yu are with the Department of Electrical and Electronic Engineering, Southern University of Science and Technology, Shenzhen, China, and also with the GaN Device Engineering Technology Research Center of Guangdong, No 1088, Xueyuan Rd., Xili, Nanshan District, Shenzhen, Guangdong, China (e-mail: yuhy@sustc.edu.cn).

Jian Zhang and Yulong Jiang are with the Department of Microelectronics, Fudan University, China. Guangnan Zhou and Guangrui Xia are with the Department of Materials Engineering, University of British Columbia, Canada.

substrates. The full process flows and the device structure are illustrated in Figure 1. After removing native oxides with a HCl: $H_2O$ (1:4) solution, La (15 nm)/Ti (20 nm)/Al (90 nm)/Ti (20 nm) (La-based) layers were deposited by a sputter (model Lab18 by Kurt J. Lesker). Then, it was followed by a lift-off process and an annealing step in $N_2$ ambient for 45 to 60 s at 830 °C. Considering that La and water can react very readily to generate hydrogen [3], we only infiltrated the samples in DI-water for 5 s after N-Methyl pyrrolidine (NMP)/iso-Propyl alcohol (IPA) lift-off to remove the IPA residues. The electrical characterizations of the Ohmic contact metal pads were performed by a circular transfer length model (CTLM) method with concentric circular metal lines of 5, 10, 15, 20, 25 and 30 µm diameters. The control devices and the control contact structures have Ti (20 nm)/Al (110 nm)/Ti (20 nm)/Au (50 nm) (Au-based) metal stacks in the S/D regions. After finishing the source and drain pads, the Ni (20 nm)/Au (50 nm) Schottky gates were deposited by e-beam evaporation, followed by a lift-off process.

### *B. Electrical and Material Characterizations*

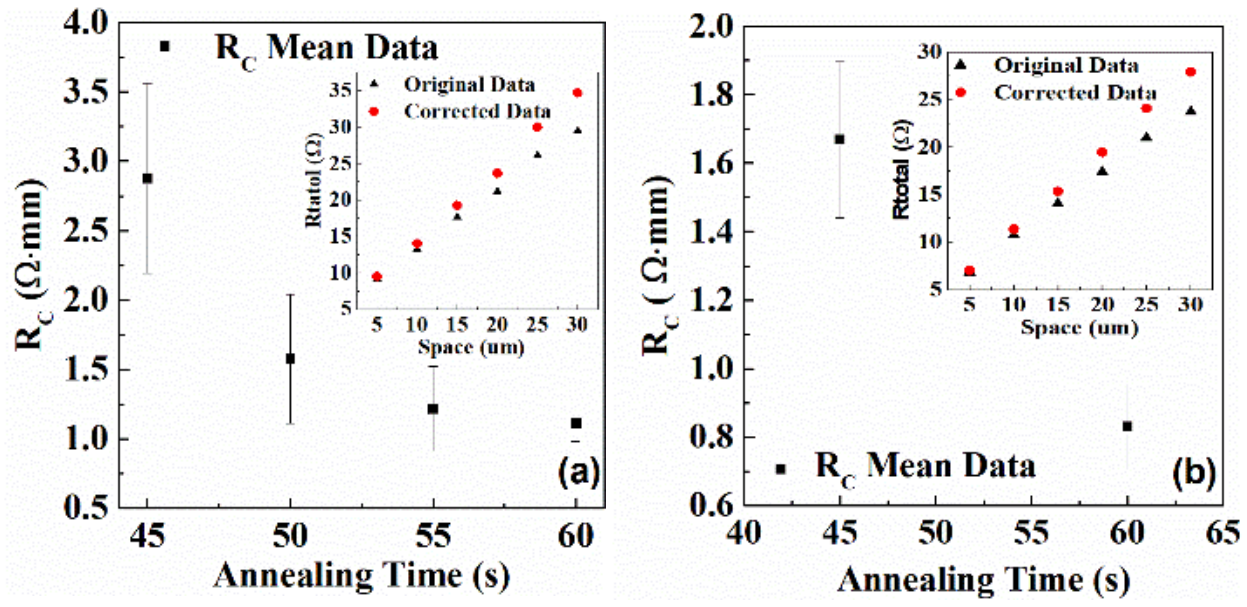

Figure 2. Rc as a function of the annealing time for (a) La-based ohmic contacts and (b) Au-based ohmic contacts.

Figure 2(a) shows the contact resistance $R_C$ as a function of the annealing time, 45 to 60 s with an interval of 5 s, of the La-based contacts to AlGaN/GaN heterostructures. The insert shows the original and corrected $R_C$ using a CTLM model for La-based contacts with 830 ºC 60 s annealing. The best achieved $R_C$ of 0.97 Ω·mm can be obtained with 60 s annealing. The correlation coefficient, which parameterizes the quality of the linear fit, of the above analysis for the AlGaN/GaN heterostructure is 0.9989. For the Au-based metallization on AlGaN/GaN HEMT structures, 60 s annealing time at 830 °C resulted in a $R_C$ of 0.72 Ω·mm, as seen Figure 2(b).

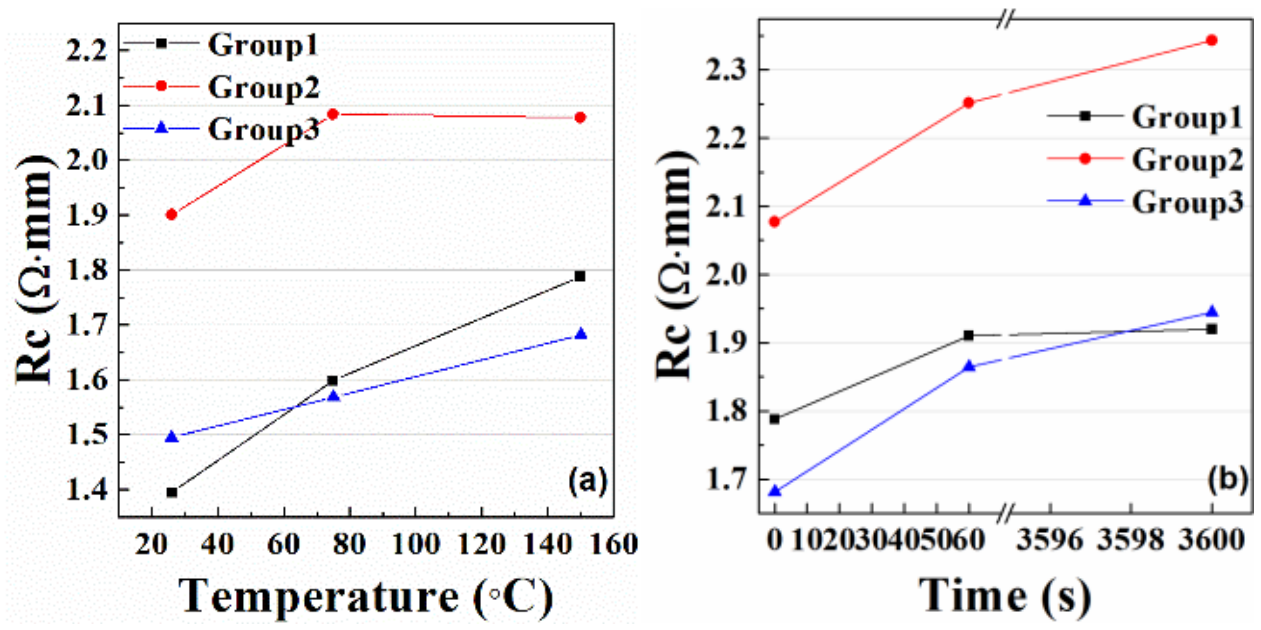

Figure 3. $R_C$ data of La-based Ohmic contacts (a) as a function of the measurement temperature from 25 to 150 °C; (b) measured at 150 °C as a function of the heating time from 0 to 3600 s.

A temperature dependent test from 25, 75 to 150 °C, was carried out to test the reliability of the La-based Ohmic contacts. Figure 3(a) and (b) show the thermal reliability testing results of three La-based Ohmic contacts in the temperature range from 25 to 150 °C. $R_C$ is plotted as a function of the temperature and time, respectively. From 25 (RT) to 150 °C, the $R_C$ value increased by 9.25% to 28.21% for three groups of devices. However, during the prolonged heating at 150 °C up to 3600 s, $R_C$ increased by 7.36% to 15.65% compared with RT. This may be related the La diffusion and segregation to the surface and the corresponding morphology change shown in Figure 5 (a) and 6.

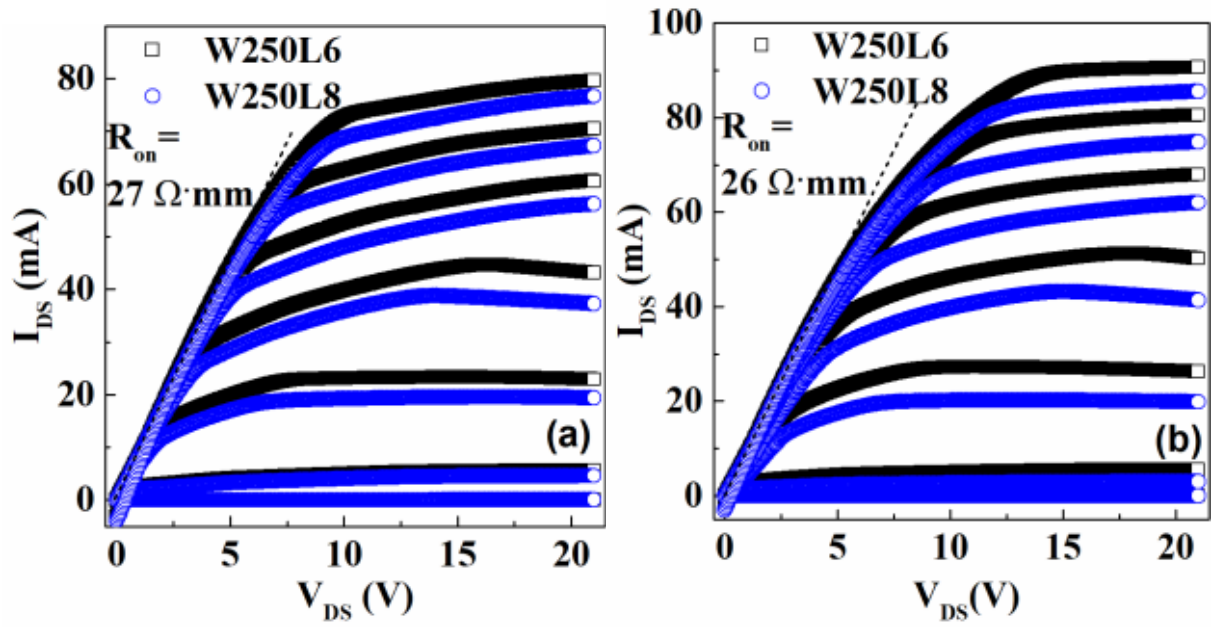

Figure 4. $I_{DS}$-$V_{DS}$ output characteristic of HEMTs with (a) La-based S/D and (b) with Au-based S/D. $V_G$ was from -4V to 2V in a step of 1V. $V_{DS}$ start from 0 V to 21V for the measurements. Width=250 µm, and Length= 6 (black squares) and 8 µm (blue circles).

Figure 4 shows the $I_{DS}$-$V_{DS}$ characteristics of the fabricated GaN transistors with (a) La-based and (b) Au-based Ohmic S/D, respectively. The obtained specific on-state resistances are 27 Ω·mm and 26 Ω·mm, respectively. The turn-on resistance of the two devices is similar but relatively large. This might be explained by 1) the devices are not passivated, resulting in a large surface state that suppresses the concentration of 2DEG; 2) the Ohmic contact resistance of 1 Ω·mm is still relatively large; and 3) Au-based Schottky gates [7].

### *C. Mechanisms of Low $R_C$ Achieved with La/Ti/Al/Ti*

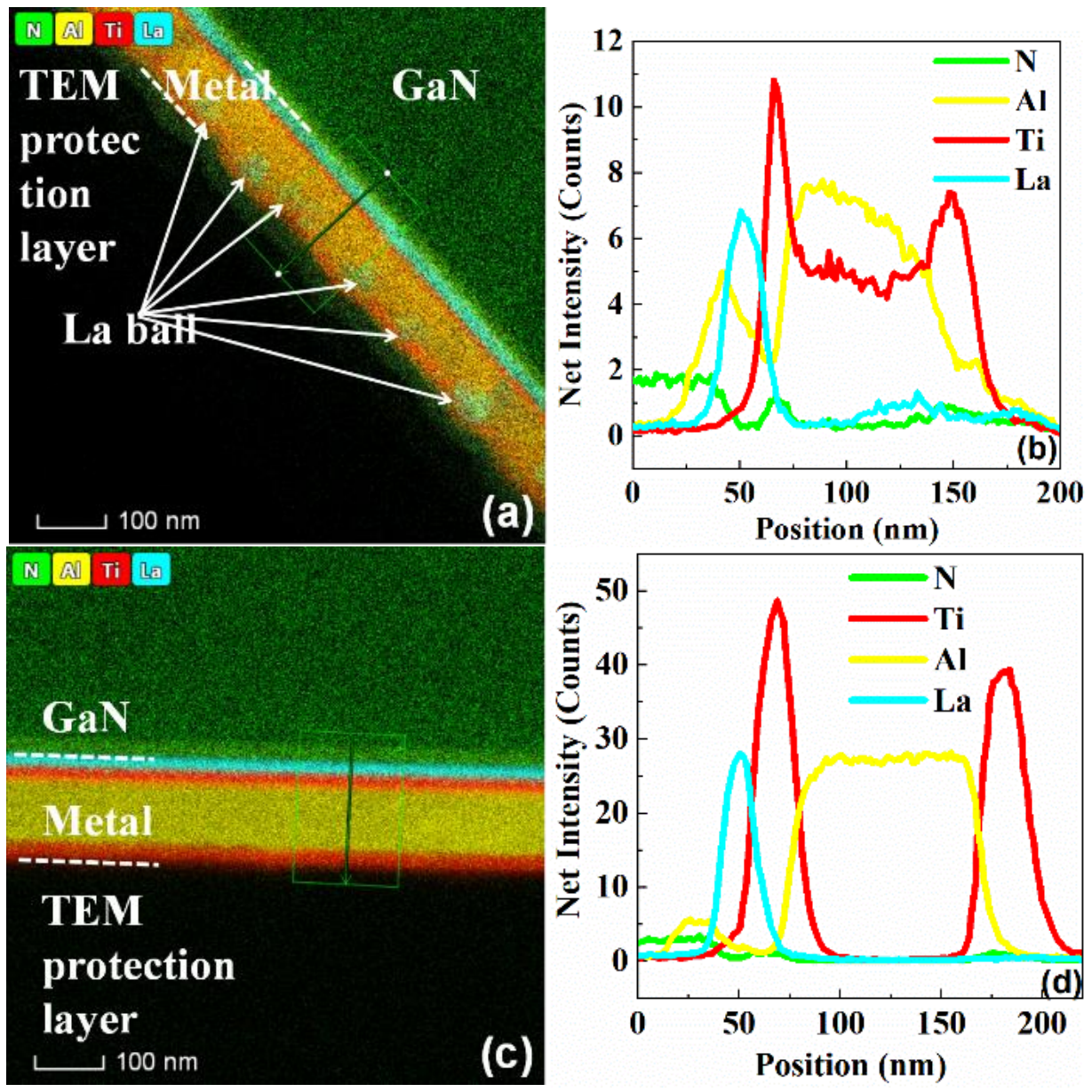

Figure 5. TEM images and the corresponding EDX diagrams of N, Al, Ti and La. (a), (b) after and (c), (d) before the annealing.

Energy dispersive X-ray spectroscopy (EDX) within a transmission electron microscope (TEM, FET Talos Model) was also used to characterize La-based (La/Ti/Al/Ti) contacts. In the TEM image in Figure 5 (a), La is in cyan. At the La/AlGaN interface, a uniform La-rich layer can be seen. Near the top of the metal stacks, spherical-shaped high La regions (LaTiAl as indicated in Figure 6) can also be observed, which suggests that La diffuses to the surface and segregates there. These spheres are likely responsible for the wavy metal surface

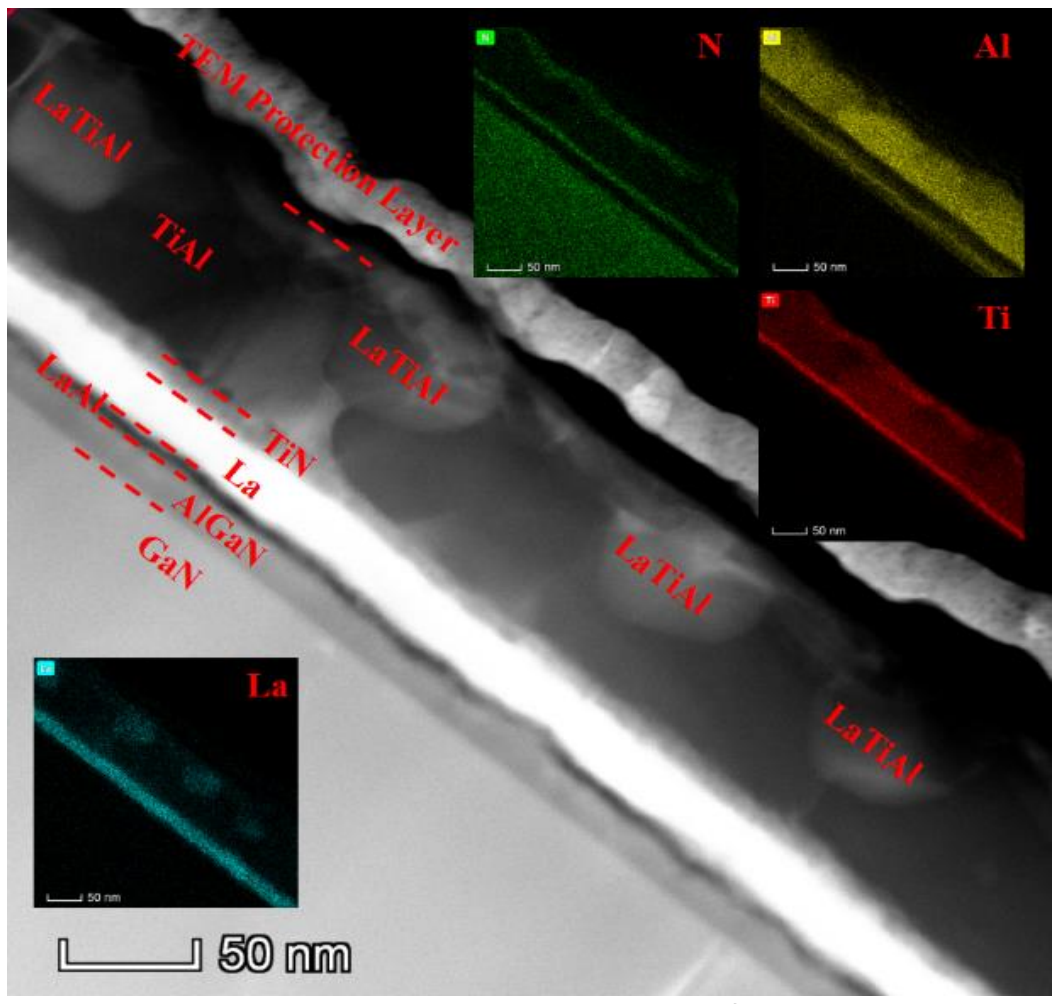

Figure 6. TEM cross-section characterization of La-based Ohmic contacts: high angle annular dark field (HAADF) micrographs of La-based contact and N/Al/Ti/La elements distribution.

after the annealing (Figure 6).

It can be seen from EDX diagram in Figure 5 (b) that after annealing at 830 °C 60s, La diffuses into the AlGaN layer, and the overlap of La and Al peaks is significantly increased compared with that before the annealing, Figure 5 (b). On the one hand, La's low work function (3.5 eV) is beneficial for reducing the barrier between the metals and GaN, as seen in Figure 7. On the other hand, a new LaAl layer formed at the metal-to-semiconductor interface after annealing, as evidenced by the HAADF TEM image seen Figure 6. The phenomenon that La diffuses into the AlGaN layer contributes to the formation of Ohmic contacts. This is similar to conventional Ti/Al/X/Au contacts, where Ti and N peaks overlap.

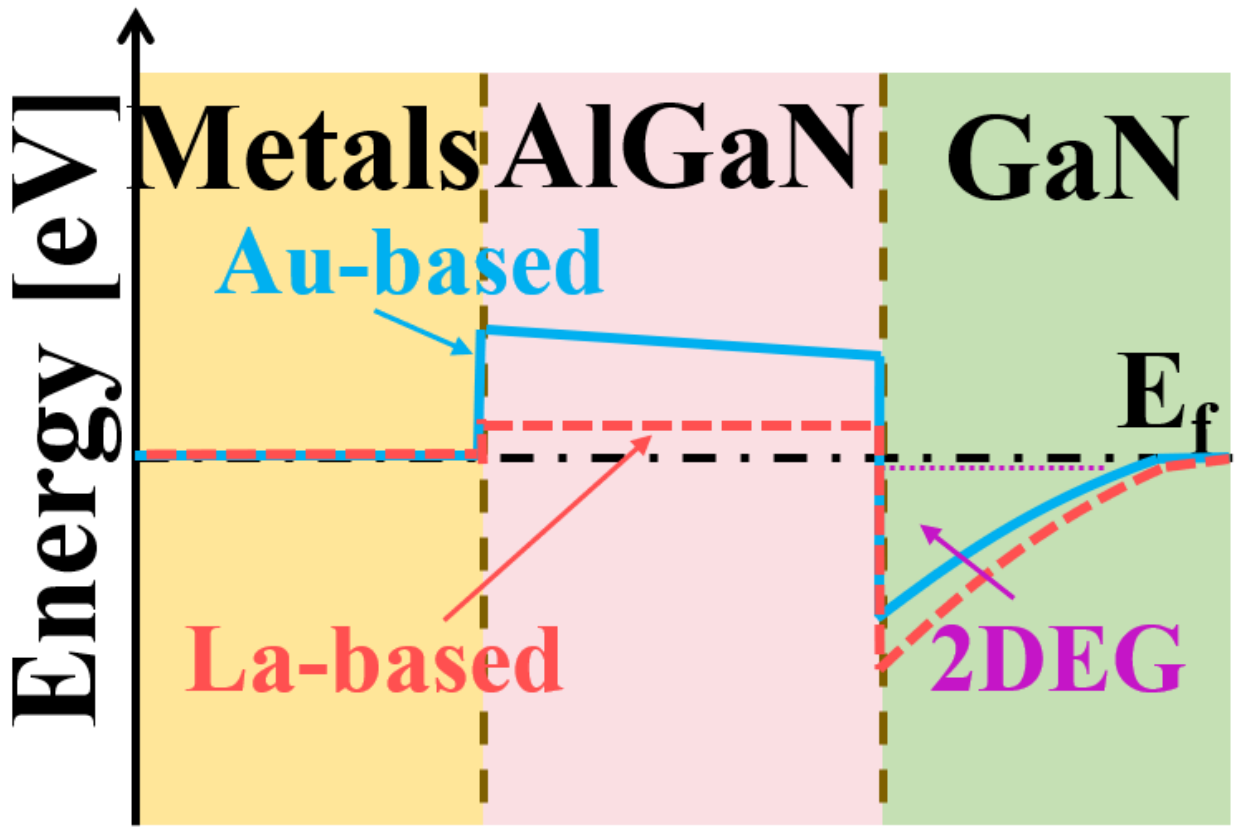

Figure 7. Band diagram for the two different Metals/AlGaN/GaN structures at equilibrium.

## III. Conclusion

In summary, La-based Ohmic contacts with non-recess process have been first made on AlGaN/GaN heteroepitaxy structures and in HEMTs. A contact resistance of 0.97 Ω.mm was achieved by the La-based metal contacts. $R_C$'s temperature dependence and reliability measurements were performed and discussed. TEM results reveal the mechanisms of the Ohmic contact formation, which combines the low work function and the new LaAl interfacial layer formation. This work demonstrates the effectiveness of La metal in the fabrication of CMOS-compatible GaN HEMTs.